\documentclass[a4paper,fleqn]{cas-dc}
\usepackage{mathrsfs}
\usepackage{siunitx}
\usepackage[latin1]{inputenc}
\usepackage{bbm}
\usepackage{graphicx}
\usepackage{graphics}
\usepackage{epsfig}
\usepackage{epstopdf}
\usepackage{xcolor}
\usepackage{mathrsfs}
\usepackage[latin1]{inputenc}
\usepackage{float}

\usepackage[numbers]{natbib}

\def\tsc#1{\csdef{#1}{\textsc{\lowercase{#1}}\xspace}}
\tsc{WGM}
\tsc{QE}
\tsc{EP}
\tsc{PMS}
\tsc{BEC}
\tsc{DE}

\begin{document}
\let\WriteBookmarks\relax
\def\floatpagepagefraction{1}
\def\textpagefraction{.001}
\shorttitle{\textit{Successful production of Solution Blow Spun YBCO+Ag complex ceramics}}
\shortauthors{AL Pessoa et~al.}

\title [mode = title]{Successful production of Solution Blow Spun YBCO+Ag complex ceramics}                      



\author[1]{AL Pessoa}[
                     orcid= 0000-0002-7949-4626]

\ead{alexsanderpessoa@yahoo.com.br}
\credit{Investigation, Data curation, Writing - Original Draft}

\address[1]{Superconductivity and Advanced Materials Group, S\~ao Paulo State University (UNESP) Campus at Ilha Solteira, Brazil}

\author[2]{MJ Raine}[
                     orcid= 0000-0001-6566-6039]
\ead{m.j.raine@durham.ac.uk}

\credit{Investigation, Data curation, Writing - Original Draft, Writing - Review & Editing, Validation}

\address[2]{Superconductivity Group, Centre for Materials Physics, Department of Physics, Durham University. DH1 3LE, UK}

\author[2]{DP Hampshire}
\ead{d.p.hampshire@durham.ac.uk}[
                     orcid= 0000-0001-8552-8514]
\credit{Conceptualization of this study, Methodology, Resources, Writing - Review & Editing, Supervision, Project administration, Funding acquisition}

\author[3]{D K Namburi}[
                     orcid= 0000-0003-3219-2708]

\ead{ndevendra@gmail.com}

\credit{Data curation, Writing - Review & Editing, Validation}

\address[3]{Department of Engineering, Trumpington Street, Cambridge CB2 1PZ, UK}

\author[3]{J H Durrell}[
                     orcid= 0000-0003-0712-3102]
\ead{jhd25@cam.ac.uk}
\credit{Writing - Review & Editing, Supervision, Validation}

\author[1]{R Zadorosny}[
                     orcid= 0000-0002-2419-2049]
\cormark[1]
\ead{rafael.zadorosny@unesp.br}

\credit{Conceptualization of this study, Methodology, Resources, Writing - Original Draft, Supervision, Project administration, Funding acquisition}

\cortext[cor1]{Corresponding author}


\begin{abstract}
YBCO fabrics composed of nanowires, produced by solution 
blow spinning (SBS) are so brittle that the Lorentz force 
produced by induced currents can be strong enough to damage 
them. On the other hand, it is known that silver addition 
improves the mechanical and flux pinning properties of 
ceramic superconductors. Thus, in this work, we show how we 
successfully obtained a polymeric precursor solution 
containing YBCO$+$Ag salts, which can be spun by the 
SBS route to produce ceramic samples. Yttrium, barium, 
copper, and silver metal acetates, and polyvinylpyrrolidone 
(PVP) (in a ratio of 5:1wt [PVP:acetates]) were dissolved 
in a solution with 61.5 wt\% of methanol, 12 wt\% of propionic 
acid, and 26.5 wt\% of ammonium hydroxide, together with 6 wt\% 
of PVP in solution. Three different amounts of silver 
(10 wt\%, 20 wt\%, and 30 wt\%) were 
used in YBa$_2$Cu$_3$O$_{7-x}$. The 
TGA characterizations revealed a lowering of crystallization and partial melting temperatures by about \SI{30}{\celsius}. 
SEM images show that after burning out the polymer, a fabric 
composed of nanowires of diameters up to \SI{380}{\nano \metre} is produced. 
However, after the sintering process at \SI{925}{\celsius} for \SI{1}{\hour}, 
the nanowires shrink into a porous-like sample. 

\end{abstract}


\begin{highlights}
\item synthesis of precursor solution with YBCO+Ag applicable to solution blow spinning
\item production of fabric-like ceramic from sol-gel route
\item shrinkage of samples due to Ag-addition
\end{highlights}


\maketitle

\section{Introduction}
\label{intro}

The main applications of superconductors are based on devices
made by low-temperature materials like NbTi \cite{1} and 
Nb$_3$Sn \cite{2}. However, since the discovery of ceramic 
high-temperature superconductors (HTS), efforts have been made 
to develop materials and devices with properties and 
forms specific to each required application. The pros of 
using HTS in turbines, generators, motors, 
magnetic shielding, and NMR/MRI are the reduced weight, 
high efficiency, compact size, low noise, high trapped fields, 
and so on \cite{3,4}. On the other hand, the cons of using 
HTS are their high production cost, high ac-losses, 
non-homogeneous trapped field distribution, and high cost 
and reliability of the cooling systems. Some of these 
issues, however, can be solved by producing materials following 
facile and low-cost routes and aiming for high values of critical current density  
$J_c$ and its homogeneous distribution along the length of the 
materials. 
Additionally, high porous superconductors, such as those  
produced by solution-blow spinning (SBS) \cite{5,6}, 
electrospinning \cite{7,8}, and in foam-like structures 
\cite{9,10}, could be used to increase cooling efficiency due to their
increased surface areas. 

Particularly in the case of SBS, the samples have a 
fabric-like structure formed by network of wires that produces a 
thin porous material. However, as can be seen by the data in 
Ref. \cite{net} the samples are very fragile; they are 
pulverized during magnetic measurements by the Lorentz force generated by the induced currents 
in the wires. Therefore, to study their superconducting properties 
in a wide range of fields and temperatures, it is necessary 
to improve their mechanical properties. 

Research works carried out on bulk (RE)BCO materials, specially in YBa$_2$Cu$_3$O$_{7-x}$ (YBCO) system, showed that both the the mechanical \cite{11,12,durrell,bz} and 
superconducting \cite{bz,13,14} properties could be significantly improved through addition of silver. Some of the other benefits of adding silver is that 
it does not chemically react with YBCO \cite{14,15,16}, 
it improves pinning sites \cite{17}, it modifies weak-links 
\cite{18,19,20}, and it enhances $J_c$ \cite{15,21}. 

A variety of silver composites have been added in YBCO 
system, such as metallic Ag \cite{15}, Ag$_2$O \cite{11}, 
and AgNO$_3$ \cite{12,21}. The samples were usually 
produced by solid-state reaction \cite{12,15,21}, 
melting-growth-like processes \cite{11}, and by electrochemical  
routes \cite{22}. In Ref. \cite{23}, a sol-gel chemical 
route was used to dope the barium site by silver in the production of YBCO pellets. It is reported that high 
concentrations of silver depreciate the superconducting 
properties but in small concentrations, it slightly enhances the critical temperature
$T_c$ and critical current density $J_c$. However, as discussed in Ref.~\cite{review}, there is 
some controversy as to the extent to which Ag can be doped into 
YBCO samples.
Also, to our best knowledge, there is no information about how 
the inclusion of silver affects the production process of 
porous samples using chemical routes, such as in SBS.

The SBS technique was first reported in Ref. \cite{24}. 
In such a technique, polymer solutions are spun by compressed 
air from an inner needle up to a collector 
\cite{24,25}. Along the working distance i.e. the space between 
the needle and the collector, the solvents have to 
evaporate, allowing the formation of stretched, thin fibers 
with diameters of the order of hundreds of nanometers. Thus, 
solutions with no water (or with very low 
water content) are crucial to this technique to ensure that the volatility of 
the solution remains sufficiently high. Apart from some silver 
composites being easily dissolved in a variety of solvents 
(including water), the synthesis of a precursor solution 
with Y, Ba, Cu, and Ag ions is greatly challenging. Here, we describe 
a sol-gel-based synthesis method for obtaining 
fabric-like samples using SBS and we present structural characterizations showing the influence of silver content on those properties.

\section{Materials and Methods}
\label{materials}

One-pot-like method \cite{one-pot} was used to synthesize the precursor 
solution, and the reagents used are shown in Table \ref{tab1}. All salts are 
heated at \SI{100}{\celsius} for about \SI{24}{\hour} before being weighed, 
ensuring that there 
is no moisture in the salts. This is particularly important as some of these salts are hydrophilic in nature. 

\begin{table*}
\centering
\begin{tabular}{l l l l }
\hline
\textbf{Reagents}  & \textbf{Chemical formula}  & \textbf{Purity (\%)}  & \textbf{Brand}\\
\hline
Yttrium acetate	& C$_6$H$_9$O$_{6.x}$H$_2$O &	99.9	&  Sigma \\
Barium acetate 	& C$_4$H$_6$BaO$_4$ &	99	& Sigma-Aldrich\\
Copper acetate	& C$_4$H$_6$CuO$_4$H$_2$O	& 99	& Sigma-Aldrich\\
Silver acetate	& C$_2$H$_3$AgO$_2$	&	99	& Sigma-Aldrich\\
Poli(vinylpyrrolidone)*	& (C$_6$H$_9$NO)$n$	&	99.99	& Sigma-Aldrich\\
Propionic acid	& C$_3$H$_6$O$_2$	&	99.5	&  Sigma-Aldrich\\
Methanol	& CH$_3$OH	&	99.8	&  Vetec \\
Ammonium hydroxide	& NH$_4$OH	&	PA	& Dinamica\\
\hline \\ 
*PVP$ = 1.300.000$ \SI{}{\gram \per \mole}
\end{tabular}
\caption{List of reagents used in the synthesis of precursor solutions.}
\label{tab1}
\end{table*}

\subsection{Sol-gel process}

\begin{figure}
	\centering
		\includegraphics[scale=.4]{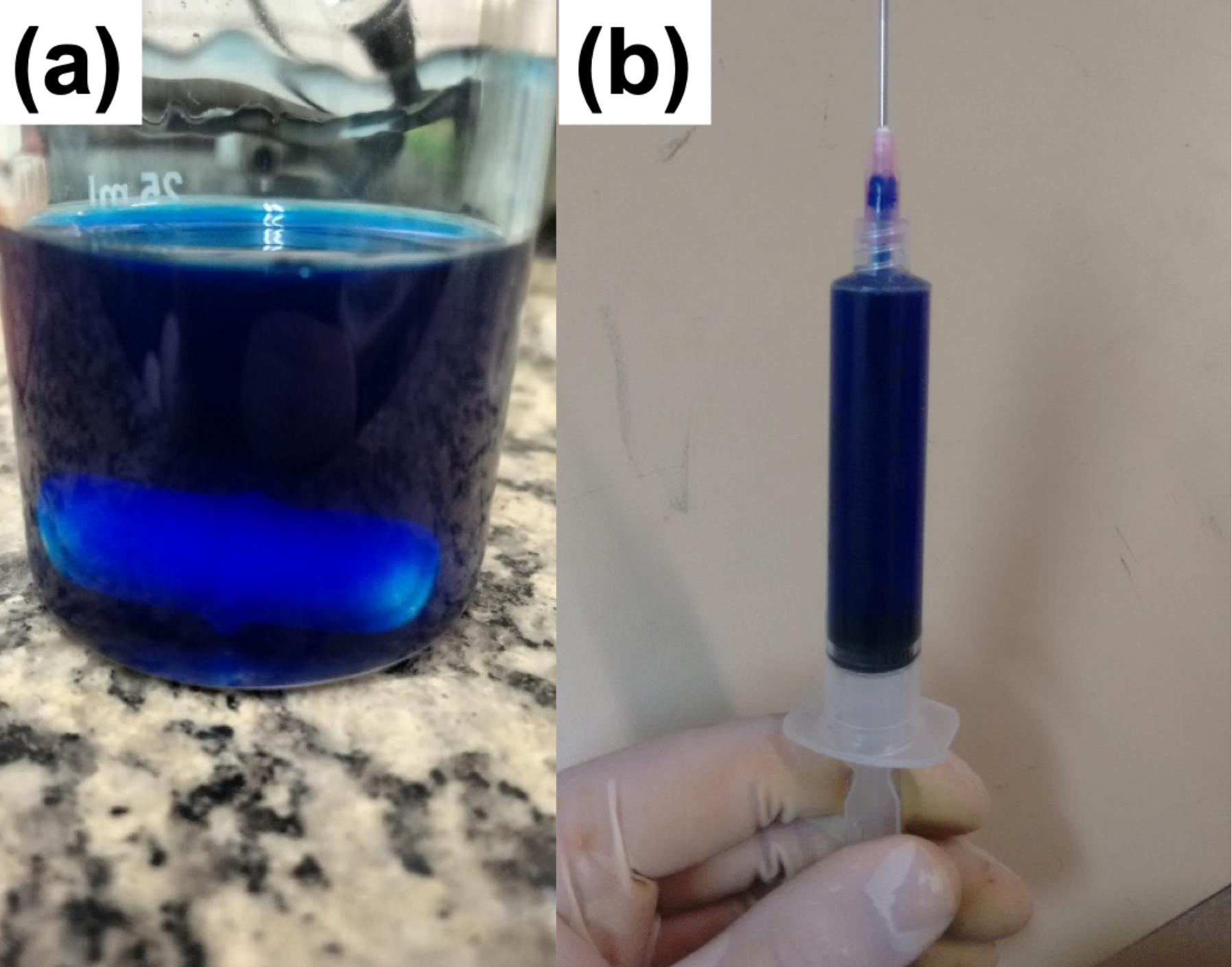}
	\caption{(a) The YAg10 precursor solution. (b) Precursor solution loaded in a \SI{10} {\milli \litre} syringe used in the SBS technique.}
	\label{fig1}
\end{figure}

The Y, Ba, and Cu acetates were stoichiometrically weighed in the molar 
ratio 1:2:3, respectively. The Ag acetate was weighed in three
concentrations namely 10 wt\%, 20 wt\%, and 30 wt\% with respect to the 
final mass of ceramic YBCO. Based on the amount of 
acetates, PVP was weighed in the ratio 5:1 (acetates:PVP). The mass 
of solvent was set to ensure that the concentration of PVP in solution was 6 wt\%. 
The acetates were placed in a vessel in the specific order Y, Ba, 
Cu, and Ag, and then propionic acid (12 wt\%), methanol (61.5 wt\%), and 
ammonium hydroxide (26.5 wt\%) were added. After five minutes of stirring, 
the PVP was added. The final precursor solution was magnetically stirred 
for \SI{24}{\hour}, with the vessel closed hermetically at room temperature (around 
\SI{28}{\celsius}). Figure ~\ref{fig1}(a) shows the stabilized YBCO precursor 
solution and in panel (b) a loaded syringe used in the SBS apparatus. 
The samples studied in the present work are labeled as YAg0, YAg10, YAg20, and YAg30 with correspondence to the concentration of Ag in YBCO as 0 wt\%, 10 wt\%, 20 wt\%, and 30 wt\%, respectively. 

\subsection{Solution-blow spinning}

A \SI{10}{\ml} syringe was connected to a 25G (diameter of \SI{0.5}{\mm}) needle. The air pressure of the compressed air cylinder was adjusted to \SI{1}{\kPa} and the working distance between the needle and the collector was set to \SI{40}{\cm}. The cylindrical collector was rotated at \SI{40} rpm and the solution within the syringe was injected into the compressed gas airflow at a rate of \SI{3}{\ml \per \hour}. A halogen light was placed above the working distance to locally heat the ejected polymer jet, evaporating the solvents prior to the jet's impact onto the rotating collector.

\subsection{Heat-treatments}

The as-collected sample was firstly heat-treated at \SI{100}{\celsius} for \SI{1}{\hour} and  
then at \SI{150}{\celsius} for another \SI{1}{\hour} with a heating rate of \SI{5}{\celsius \per \minute}. 
Some portions of that 
sample were then used to obtain SEM images. The polymer decomposition was 
carried out at \SI{600}{\celsius} for \SI{3}{\hour} ramping the temperature up and down at a rate 
of \SI{1}{\celsius \per \minute}. Some pieces of the sample at that stage were also used to make SEM 
analysis. The synthesis was carried out in a tube furnace. 
The heat-treatment consisted of increasing the temperature from room temperature to \SI{820}{\celsius} at 
\SI{3}{\celsius \per \minute} and dwelling for \SI{14}{\hour}. 
Durign the heating process, flowing O$_2$ was turned on at \SI{500}{\celsius}. 
After that dwell period, the temperature was increased at \SI{1}{\celsius \per \minute} to 
\SI{925}{\celsius} and this was held for \SI{1}{\hour}. Then, also at \SI{1}{\celsius \per \minute}, 
the temperature 
was decreased to \SI{725}{\celsius} for \SI{6}{\hour}; then at \SI{3}{\celsius \per \minute} to \SI{450}{\celsius} for \SI{24}{\hour}. 
Finally, the O$_2$ gas flow was turned off and the temperature was decreased to room temperature  
at \SI{3}{\celsius \per \minute}.

\subsection{Characterizations}
\label{charact}

Thermogravimetric measurements were carried out on the samples YAg0 and YAg10 (obtained after a heat-treatment at \SI{600}{\celsius}), employing TA Instrument, model SDT-Q600. 
Measurements were carried out under flowing compressed air at a rate of 
\SI{100}{\milli \litre \per \minute} and the temperature was increased from \SI{25}{\celsius} to 
\SI{1000}{\celsius} at a rate of \SI{10}{\celsius \per \minute}. 
For the scanning electron microscopy (SEM) measurements, 
an  EVO LS15 Zeiss operated at \SI{20}{\kilo \volt} was used. For that, 
the samples were attached in an aluminum sample holder with carbon tape, 
and gold was sputtered on their surface for \SI{2}{\minute} (\SI{5}{\nano \metre} average thickness) 
using a QUORUM Model Q150T E.
The diameter distribution was measured using a randomly selected set of 100 wires and 
the free software package ImageJ.
The x-ray analysis (XRD) was performed in a Shimadzu XDR-6000 
diffractometer with CuK$\alpha$ radiation (wavelength: \SI{1.5418}{\angstrom}). 
The displacement ranged from 
$2\theta=$ \ang{5} to \ang{60} at a scan rate of 
\SI{1}{\degree \per \minute} and measuring in steps of \SI{0.02}{\degree}. 

\begin{figure}
	\centering
		\includegraphics[scale=.16]{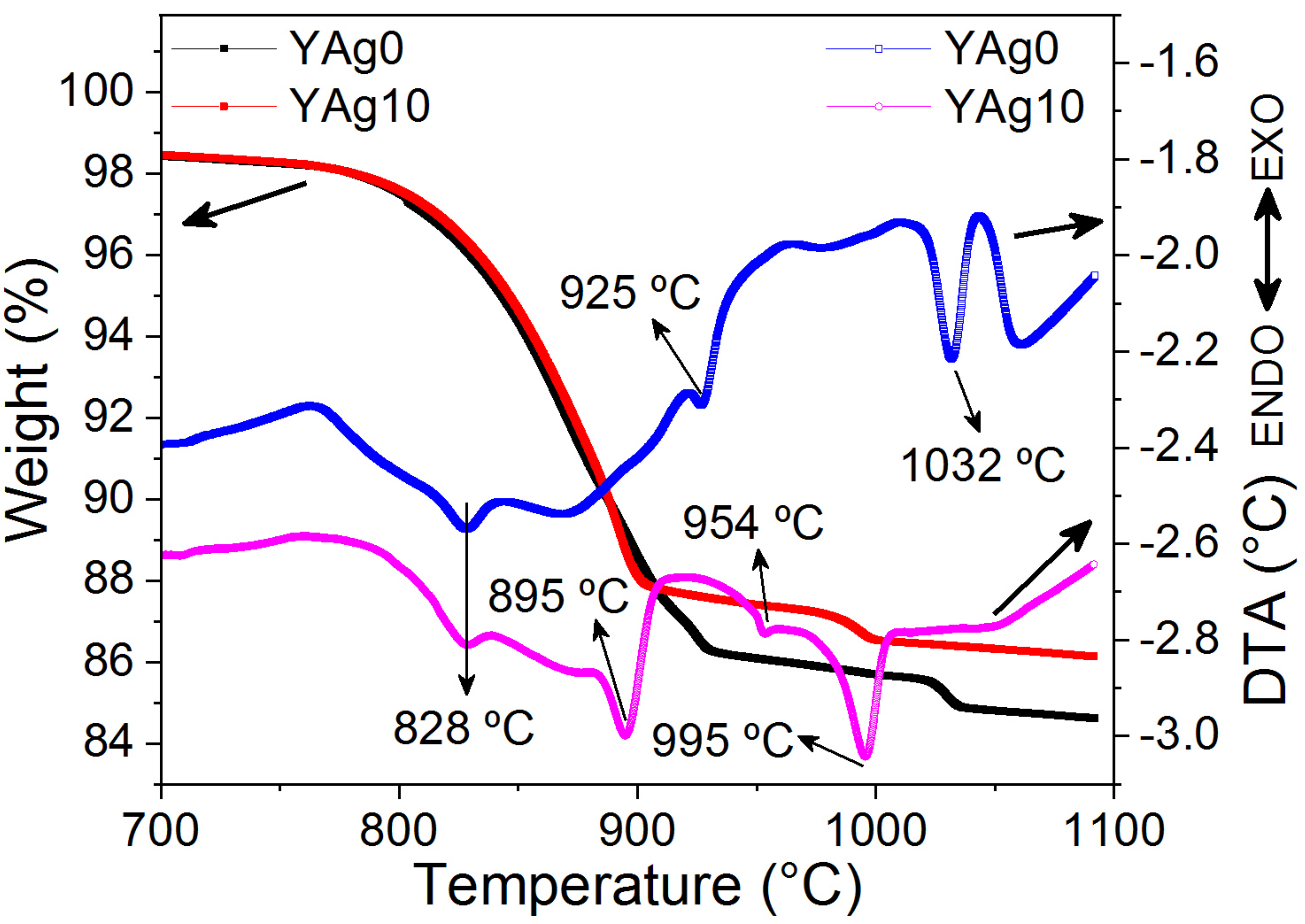}
	\caption{TG and DTA of samples YAg0 and YAg10 carried out after a heat-treatment at \SI{600}{\celsius}. The YAg10 lost less mass due to the silver addition. From DTA curves, it is noted that the silver addition shifted the YBCO crystallization peak from \SI{925}{\celsius} (YAg0) to \SI{895}{\celsius} (YAg10). The partial melting peak is also shifted from \SI{1032}{\celsius} for YAg0 to \SI{995}{\celsius} for YAg10. The peak at \SI{954}{\celsius} is related with the melting of metallic silver.}
	\label{fig2}
\end{figure}

\section{Results and Discussions}
\label{results}

Thermal analysis was carried out on two samples YAg0 and 
YAg10, after both were heat treated at \SI{650}{\celsius}. 
About \SI{15}{\milli \gram} of each sample was used for the measurement and the data obtained is 
shown in Figure ~\ref{fig2}. Both the samples exhibited a similar 
weight loss, as can be seen from the curves in Figure ~\ref{fig2}. The mass lost between \SI{25}{\celsius} and \SI{700}{\celsius} 
(not shown in Figure ~\ref{fig2}),
was 1.5 \%, and can be associated to the evaporation of water 
adsorbed in the surface of the samples from the atmosphere or even some organic groups 
remaining after the heat-treatment. It is also observed that YAg10 lost about 1.1\% 
less mass than YAg0 due to silver addition. It is not shown here, but it is worth 
pointing out that the PVP degradation occurs between \SI{400}{\celsius} and 
\SI{550}{\celsius} \cite{8,yuh,duarte}.  

The DTA curves of YAg0 and YAg10 are quite distinct. Both samples present an 
endothermic peak at \SI{828}{\celsius} which can be associated with the reaction 
between Y$_2$Cu$_2$O$_5$ and BaCuO$_2$ forming YBCO \cite{pathak,kalanda}. 
The endothermic peak at \SI{925}{\celsius} for YAg0 can be associated with the YBCO
crystallization, and it was the temperature chosen to be applied in all samples
presented in this study. However, it can be noted that the YBCO begins to crystallize 
at about \SI{895}{\celsius} for YAg10, which means that the silver addition reduced the  the crystallization temperature 
by \SI{30}{\celsius} (or 3\%). The peak 
at \SI{954}{\celsius} presented by YAg10 is associated with the melting of 
metallic silver \cite{kohayashi}. 
On the other hand, the endothermic peak at \SI{1031}{\celsius} for YAg0 is due to 
a partial melting of 
YBCO and such a peak is shifted to \SI{995}{\celsius} (or 3.5\%) for YAg10, showing 
that the silver addition also decreases this temperature \cite{durrell,nakamura,cai}.

\begin{figure}
	\centering
		\includegraphics[scale=.45]{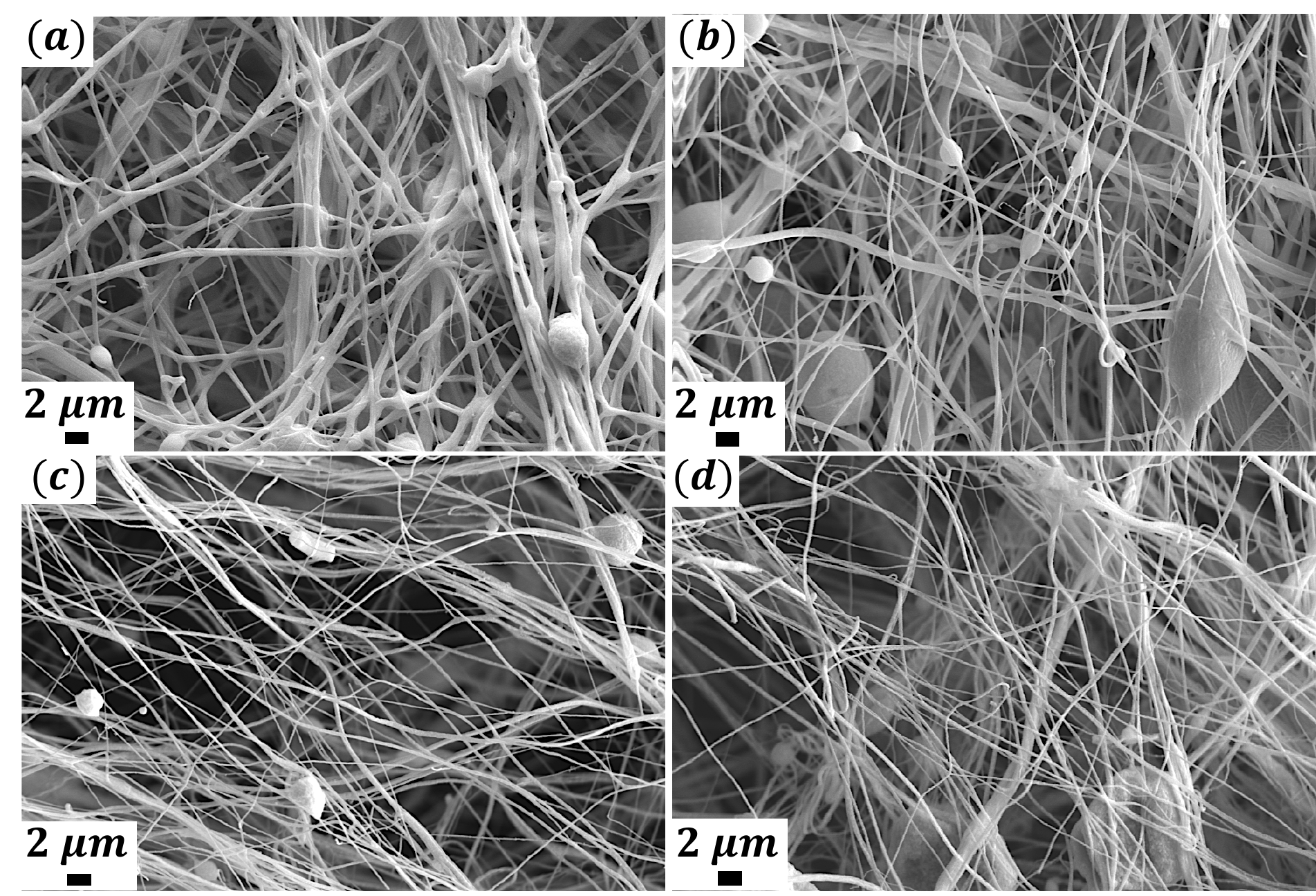}
	\caption{SEM images of samples (a) YAg0, (b) YAg10, (c) YAg20, and (d) YAg30 after calcination at \SI{600}{\celsius}. At this step, all samples have the fabric-like morphology with beads distributed along their lengths.}
	\label{fig3}
\end{figure}

Based on the thermogravimetric analysis of Figure ~\ref{fig2} and 
on the literature \cite{5,8,yuh,duarte}, the samples were firstly 
heat-treated at \SI{600}{\celsius} to ensure the total decomposition of the PVP. 
Figure ~\ref{fig3} shows SEM images of the produced samples. All of 
which present a fabric-like structure with randomly entangled wires, 
however, it can be seen that beads are distributed along those wires. This is probably 
due to water from the acid$-$base reaction in the precursor 
solution synthesis. Besides that, the wires are long and smooth.
Table ~\ref{tab2} shows the average diameter ($d_{av}$) of the samples. Wires in the size range \SI{180}{\nano \metre} to \SI{233}{\nano \metre} were found in the samples. No clear relationship was found between $d_{av}$ and the content of silver present in the system.  

\begin{table}
\centering
\begin{tabular}{l l l }
\hline
\textbf{Sample}  & \textbf{$d_{av}$ (nm)}  & \textbf{Deviation (nm)}\\
\hline
YAg0 & 233 & 77	\\
YAg10 & 180 & 68 \\
YAg20 & 206	& 76 \\
YAg30 & 191	& 63 \\
\hline \\ 
\end{tabular}
\caption{Average diameters of the samples heat-treated at $600 ^o$ with their respective deviations.}
\label{tab2}
\end{table}

\begin{figure}
	\centering
		\includegraphics[scale=.45]{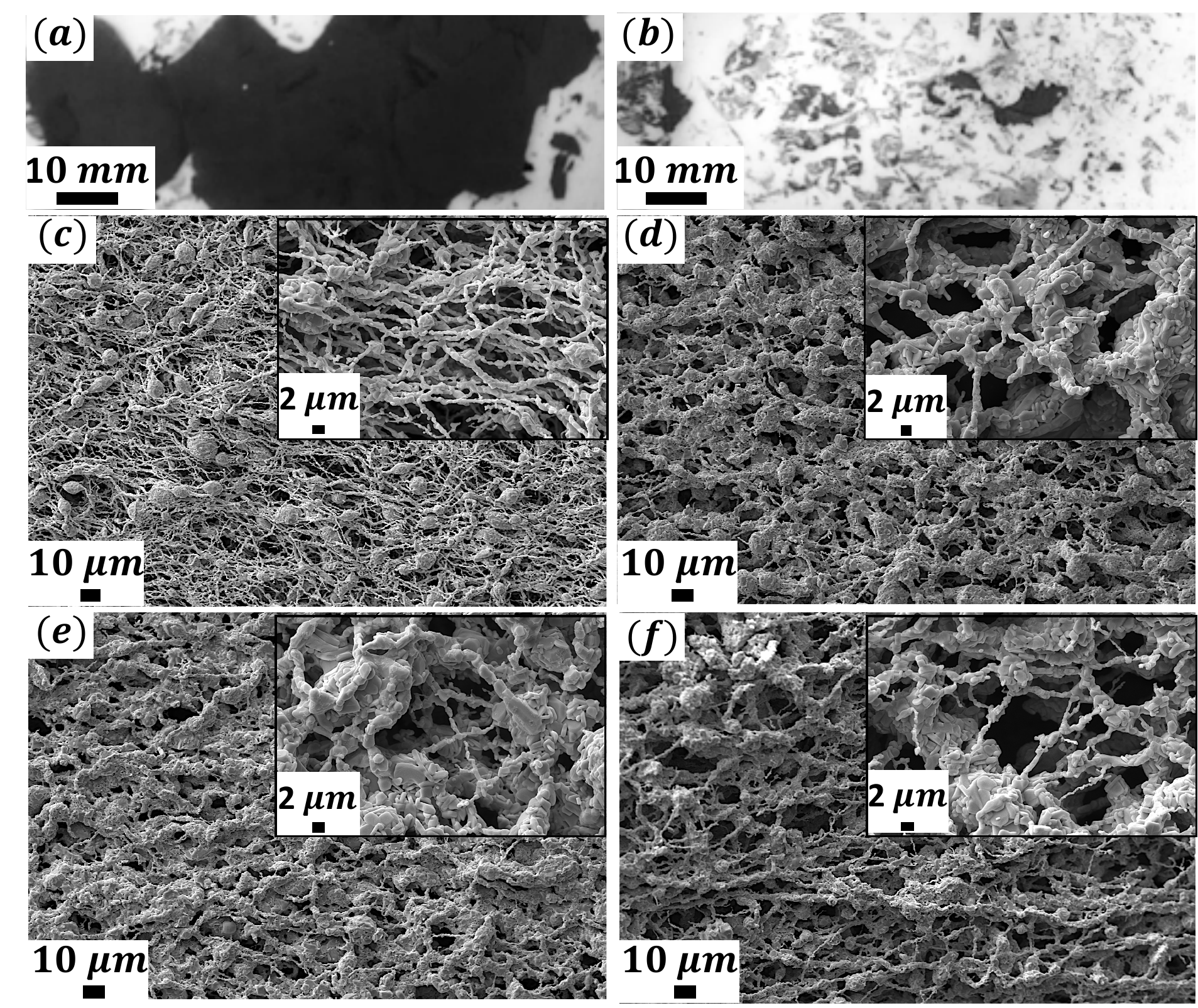}
	\caption{(a) Image of the YBCO-Ag after heat-treatment at \SI{600}{\celsius}, and panel (b) shows the visible shrinkage of the sample after sintering at \SI{925}{\celsius}. (c) SEM images of YAg0 sample showing the formation of the wires network structure. From (d) to (f) are the SEM images of the YAg10, YAg20, and YAg30, respectively, showing a porous-like structure due to shrinkage after the sintering process.}
	\label{fig4}
\end{figure}

After the sintering process at \SI{925}{\celsius}, the silver samples shrank, 
producing a granular porous-like structure. 
The shrinkage of the samples is shown in Figure ~\ref{fig4} 
(a) and (b). The scale bars within those images are an approximation for 
comparison purposes. Figure ~\ref{fig4}(c) shows that, 
while the Ag-free sample YAg0 maintains its fiber-like structure, 
the Ag$-$added samples showed considerable enhancement in density, as shown in Figure ~\ref{fig4} 
from panels (d) to (f). With the wires closer to 
each other, the grains begin to coalesce during the sintering process and 
the samples acquire a porous-like morphology.
Some works report that the heat-treatment temperatures can be reduced with 
Ag addition in YBCO bulks \cite{durrell,nakamura,cai} due to enhanced
heat diffusion. In the case of
the present work, the presence of Ag facilitates improved heat-diffusion 
between the ceramic grains, decreasing the sintering temperature of the samples.

\begin{figure}
	\centering
		\includegraphics[scale=.16]{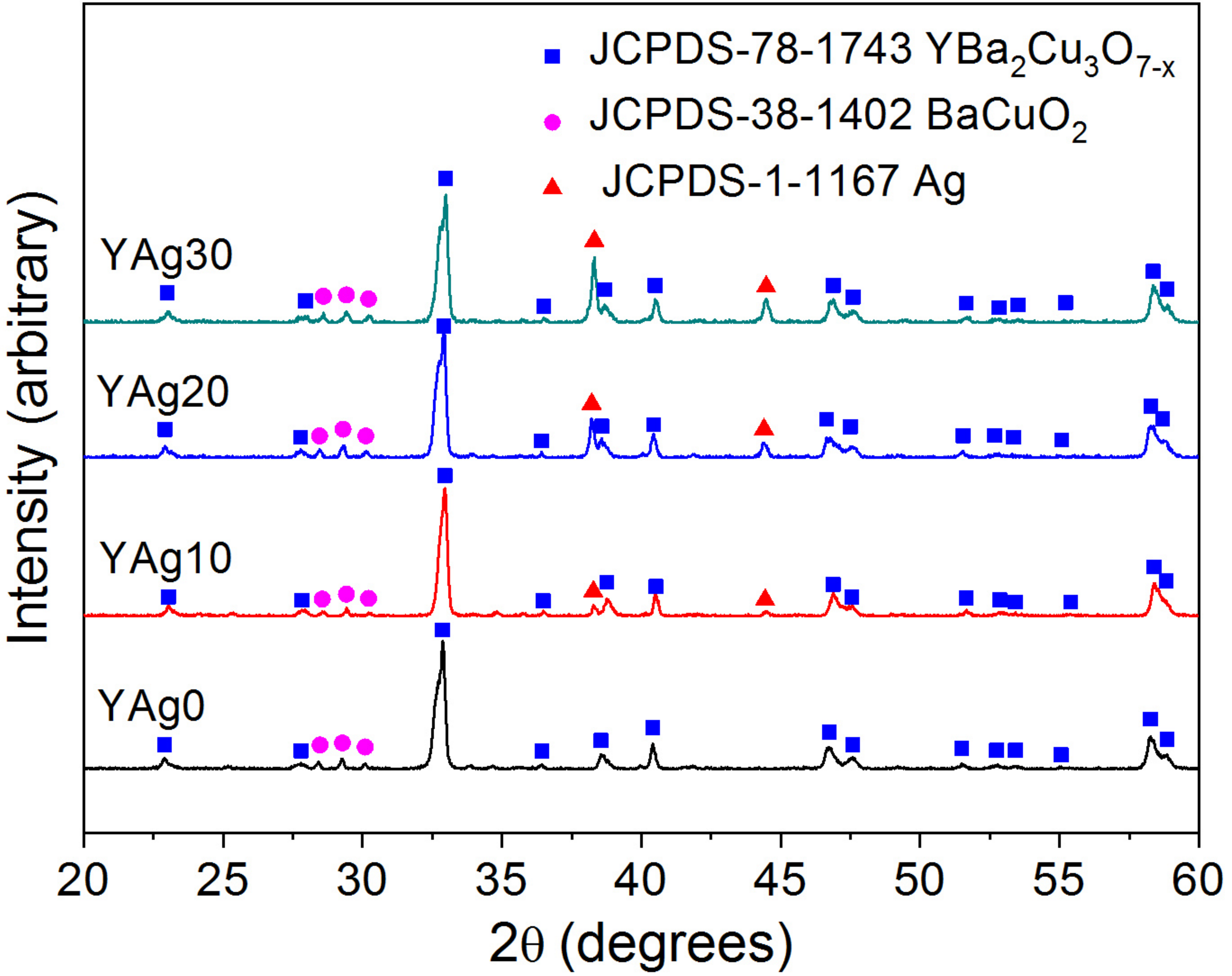}
	\caption{XRD diffractograms of the produced samples. It can be seen that BaCuO$_2$ is present in all the samples. The main YBCO peak is at 
	$2\theta=$ \SI{32.88}{\degree}, \SI{32.96}{\degree} \SI{32.92}{\degree}, and \SI{32.98}{\degree}, for YAg0, YAg10, 
	YAg20, and YAg30, respectively. The metallic silver peaks are those 
	ones at \SI{38.28}{\degree} and \SI{44.5}{\degree} for YAg10, \SI{38.22}{\degree} and 
	\SI{44.4}{\degree}
	for YAg20, and \SI{38.3}{\degree} and \SI{44.48}{\degree} for YAg30.}
	\label{fig5}
\end{figure}

Figure ~\ref{fig5} shows XRD diffractograms of all the samples currently studied where it 
can be seen that the BaCuO$_2$ phase is present within each of them. 
Since samples that were produced using PVP of \SI{360,000}{\gram \per \mole} instead of 
\SI{1300,000}{\gram \per \mole} contain a pure 
phase \cite{5}, we believe that a longer polymer chain could influence the 
ceramic phase formation. Such a study will be published in future. As the 
silver content increases, the intensity of the silver peaks (at around 
\SI{38}{\degree} and \SI{49}{\degree}) also increases. The most intense YBCO peak position 
shifts with the silver addition, being at $2\theta=$ 
\SI{32.88}{\degree}, \SI{32.96}{\degree}, \SI{32.92}{\degree}, and \SI{32.98}{\degree}, for YAg0, YAg10, 
YAg20, and YAg30, respectively. This can indicate that there is some 
saturation for silver doping above which metallic silver begins to form 
along the sample \cite{review}. The peaks around $2\theta=$ \SI{38.3}{\degree} and 
\SI{44.5}{\degree} were identified as characteristic of metallic silver, and their 
intensity increases with increasing Ag content.

\section{Conclusions}
\label{final}

In the present work we report the synthesis of YBCO-Ag nanowires via solution blow spinning SBS techniue. This approach is based on an acetate chemical route 
where yttrium, barium, copper and silver acetates were dissolved in a solution with 
propionic acid (12 wt\%), methanol (61.5 wt\%), and 
ammonium hydroxide (26.5 wt\%). The silver was added in amounts of 10 wt\%, 20 wt\%, and 30 wt\%. 
Thermogravimetric analysis show that the addition of silver decreases both the YBCO 
crystallization and the partial melting temperatures by \SI{30}{\celsius}. Both DTA and XRD 
characterizations showed the presence of metallic silver. Another interesting influence of 
silver in such complex ceramics is the huge densification of the samples sintered at 
\SI{925}{\celsius} for one hour. SEM images show that the fabric-like morphology of 
the samples heat-treated at \SI{600}{\celsius} is lost with the sintering process, for
which a porous-like morphology takes place due to a shrinkage of the ceramic wire network. 
Thus, the routine described in this study could be used to produce high density 
bulk superconductors for use in applications such as flywheels, trapped magnets, motors and 
generators.

\section{Acknowledgements}
\label{thanks}
A. L. Pessoa, R. Zadorosny, M. Raine and D. P. Hampshire acknowledge the 
Brazilian agencies S\~ao Paulo Research Foundation (FAPESP, grant 
2017/50382-8), Coordena\c c\~ao de Aperfei\c coamento de Pessoal de 
N\'ivel Superior (CAPES) $-$ Finance Code 001, and National Council 
of Scientific and Technologic Development (CNPq, grant 302564/2018-7). We also 
thanks Prof. Agda E. S. Albas and Silvio R. Teixeira, from UNESP-Presidente Prudente, for 
the thermogravimetric measurements.


\end{document}